\documentclass[prl,twocolumn,showpacs,amsmath,amssymb]{revtex4-1}

\usepackage{graphicx,amsmath,slashbox,bm, braket, txfonts, color}

\begin{document}

\title{Resonantly Tunable Majorana Polariton in a Microwave Cavity}

\author{Mircea Trif}
\author{Yaroslav Tserkovnyak}
\affiliation{Department of Physics and Astronomy, University of California, Los Angeles, California 90095, USA}

\date{\today}

\begin{abstract}
We study the spectrum  of a one-dimensional Kitaev chain placed in a microwave cavity. In the off-resonant regime, the frequency shift of the cavity can be used to identify the topological phase transition of the coupled system. In the resonant regime, the topology of the system can be controlled via the microwave cavity occupation and, moreover, for a large number of photons (classical limit), the physics becomes similar to  that of periodically-driven systems (Floquet insulators). We also analyze numerically a finite chain and show the existence of a degenerate subspace in the presence of the cavity that can be interpreted as a \textit{Majorana polariton.} 
\end{abstract}


\maketitle

{\it Introduction.}|Majorana fermions, or half-fermions, have recently attracted tremendous attention as building blocks of a topological quantum computer \cite{Kitaev:2001qf,Freedman:2003kx,Nayak:2008oq,Fu:2008cr}. They are predicted to emerge as excitations in several solid-state systems, such as genuine two-dimensional (2D) $p$-wave superconductors \cite{Sato:2009zr}, or induced via superconducting proximity effects in topological insulators \cite{Fu:2008cr,Linder:2010dq,Wimmer:2010kl} or one-dimensional (1D) wires \cite{Sau:2010vn,Oreg:2010bh,Lutchyn:2010ve,Alicea:2010ys,SuhasLoss:2011} as boundary zero modes. They obey non-Abelian statistics, both in 2D \cite{Ivanov:2001nx} and 1D \cite{Alicea:2011tg}, allowing for implementation of certain (nonuniversal) gate operations required in quantum-computational schemes via braiding of the Majorana fermions.  Moreover, due to their highly nonlocal character, the qubits built out of Majorana fermions are insensitive to  local parity-conserving perturbations, thus making them potentially robust against physical noise errors \cite{Freedman:2003kx,Cheng:2011fv,Goldstein:2011dz}. 

The existence of Majorana states, however, has been shown not only for static systems, but also for driven systems, as zero modes in the so-called Floquet Hamiltonians \cite{Lindner:2011hc,Kitagawa:2010ij,Jiang:2011bs}, or even, more recently, in dissipative systems \cite{Zoller:2011rr}. These proposals are based on the idea that driven systems can have different topology from their static parents, albeit these are  not actual ground states. The ground-state perspective can, however, be used even for the driven models, if instead of a classical driving one considers  the driving as having its own quantum dynamics, such as in a high-$Q$ electromagnetic cavity. In particular, 1D microwave cavities have been proven extremely successful in reaching the so-called strong-coupling regime between photons and different types of qubits \cite{IAB+99,WSB+04,BHW+04,BI06,TGL08,I09,Jelezko:2011rr,Arzhang:2011rr,Arzhang:2011pi}, with very large $Q$ factors and a high degree of control. Moreover, strong coupling between atomic gases (e.g., in a Bose-Einstein condensate) and optical cavities have been achieved with a high degree of coherence \cite{esslinger:2010}. In this paper, we combine cavity QED with Majorana physics in a 1D lattice model. We analyze the spectrum of a $p$-wave lattice superconductor (Kitaev chain)  coupled to a microwave cavity and examine the  topology of the combined system. The spectrum is studied using  Dicke-like Hamiltonian, in both the off- and on-resonant regimes.   

\begin{figure}[t]
\begin{center}
\includegraphics[width=0.9\linewidth]{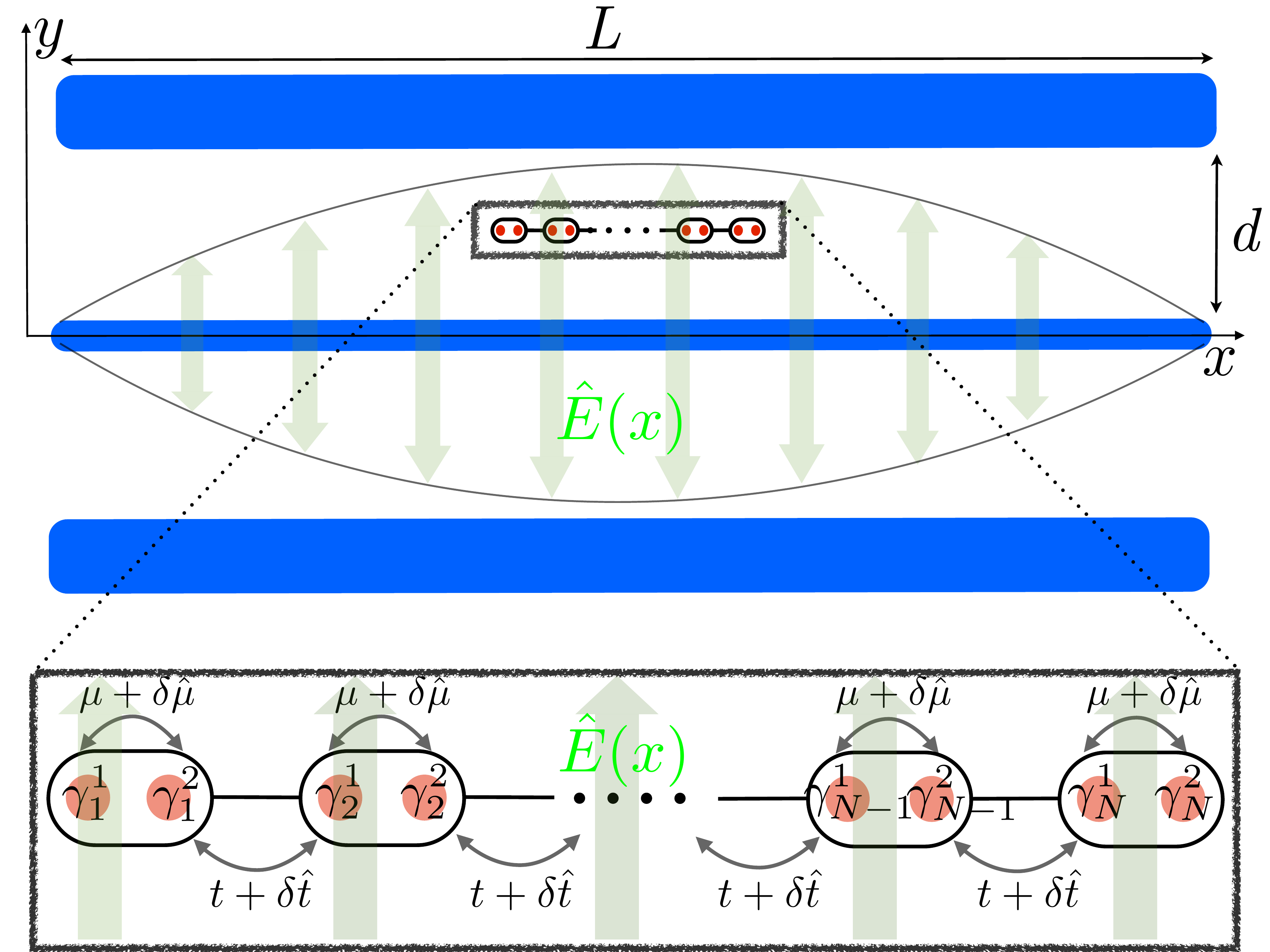}
\caption{A sketch of the system. In blue is the superconducting microwave cavity, which supports quantized electromagnetic modes. The Majorana chain $\gamma_{1,2}^1\dots\gamma_N^{1,2}$ is depicted inside the cavity with  the relevant parameters renormalized due to the coupling to the cavity: Chemical potential $\mu\rightarrow\mu+\delta\hat{\mu}$ and hopping amplitude $t\rightarrow t+\delta\hat{t}$ (with the hats denoting fluctuating quantum contributions).}
\label{sketch_system}
\end{center}
\end{figure} 

{\it The model.}|In Fig.~\ref{sketch_system}, we show a schematic of the system under consideration: a Kitaev chain inserted in a 1D microwave cavity.  The physical sites are depicted by black ovals, while the constituent Majoranas are shown inside by solid red circles. The hopping parameter $t$ and the chemical potential $\mu$ can be modified by the (quantum) cavity electric field $\hat{\bm{E}}(x)$, engendering a physical coupling between electrons and photons. The associated wavelength is assumed to be much larger than the 1D chain, so that the electric field  has no spatial dependence along the latter.

The total Hamiltonian of the system, $\mathcal{H}=H_{\rm 1D}+H_{\rm int}+H_{\rm ph}$, consists of
\begin{align} 
H_{\rm 1D}&=-\sum_{j=1}^{N-1}\left(tc^{\dagger}_{j+1}c_j+\Delta c^{\dagger}_{j+1}c^{\dagger}_j\right)+{\rm H.c.}-\mu\sum_{j=1}^Nc_j^{\dagger}c_j\,,\nonumber\\
H_{\rm int}&=\left(\beta\sum_{j=1}^Nc_{j+1}^{\dagger}c_j+{\rm H.c.}+\alpha\sum_{j=1}^Nc_j^{\dagger}c_j\right)\left(a^{\dagger}+a\right)\,,
\label{Hamiltonian}
\end{align}
and $H_{\rm ph}=\omega\,a^{\dagger}a$, where $t$ is the hopping parameter, $\Delta$ is the $p$-wave pairing potential, $\mu$ is the chemical potential, $\alpha$ is the electron-photon coupling that shifts the chemical potential, and $\beta$ accounts for changes in the tunneling Hamiltonian in the presence of photons. While the hopping change in electric field appears naturally, the modification of the chemical potential is unambiguous only in the presence of a superconducting reservoir and is associated with supercurrents flowing from and into the nanowire on a time scale determined by the strength of the associated Josephson coupling. We are thus tacitly assuming that such Josephson oscillations are much faster than the relevant dynamics in the cavity, so that the effective pairing $\Delta$ is essentially anchored at a constant value. For simplicity, we also assume that $\Delta$ is unaffected by the photon field. $a$ ($a^{\dagger}$) stands for the photon creation (annihilation) operator and $\omega$ is the frequency of the corresponding photon mode (setting $\hbar=1$ throughout). The electric field inside the cavity (correspondng to its fundamental harmonic) is $\hat{\bm{E}}(x)=\hat{\bm{e}}_y\sqrt{\omega/Ld^2c}\sin{(\pi z/L)}(a^{\dagger}+a)$ \cite{BHW+04}, with $L$ being the cavity length, $d$ is the distance between the center conductor and the ground conductors (see Fig.~\ref{sketch_system}), and $c$ is the capacitance per unit length. Note that the maximum of the electric field is at $z=L/2$, where we suppose to position the wire.

{\it Isolated wire.}|We begin by recapping some of the known results on the Majorana physics in 1D $p$-wave superconductors. To keep the ideas transparent (but without loss of generality), we assume in this section real $t=\Delta>0$ and $\mu=0$.
Performing the substitutions $\gamma_{j}^{1}=c^{\dagger}_j+c_j$  and $\gamma_j^2=i(c_j-c_j^{\dagger})$, in terms of Majorana operators, $(\gamma_{j}^{i})^{\dagger}=\gamma_j^i$, with $i=1,2$,  we can write the 1D wire Hamiltonian simply as $H_{\rm 1D}=-it\sum_{j=1}^{N-1}\gamma_{j}^2\gamma_{j+1}^1$, with  $\gamma_{1}^1$ and $\gamma_{N}^2$ dropping out entirely. This guarantees a degenerate ground state, which can be viewed as a qubit indexed by $\sigma_z\equiv (c_F^{\dagger}c_F-1/2)$. $c_F=(\gamma_{1}^1+i\gamma_N^2)/2$ here is the {\it nonlocal} fermionic operator defining the corresponding zero-mode quasiparticle.

The existence of the Majorana end modes can be identified using bulk properties only, as they are  a consequence of  the topology of the Brillouin zone. Using periodic boundary conditions allows us to switch to the reciprocal space using  $c_j=\sum_kc_k\exp{(-ikj)}/\sqrt{N}$ to write $\mathcal H_{\rm 1D}=\sum_kH_{\rm BdG}(k)$, where 
\begin{equation}
H_{\rm BdG}(k)=-\left(t\cos{k}+\mu\right)\tau_k^z-t\sin{k}\,\tau_k^x
\label{BogoGen}
\end{equation}
is the Bogoliubov-De Gennes Hamiltonian (restoring a finite $\mu$), and the pseudospin $\bm{\tau}_k=(\tau_k^x,\tau_k^y,\tau_k^z)$, expressed  in terms of the Pauli matrices, acts on the particle-hole basis $(c_k,c_{-k}^{\dagger})$. One can diagonalize this Hamiltonian $\widetilde{H}_{\rm BdG}(k)=U^\dagger(k)H_{\rm BdG}(k)U(k)$ by $U(k)=\exp{(i\theta_k\tau_k^x)}$, which gives $\widetilde{H}_{\rm BdG}=\epsilon_k\tau_k^z$, with $\epsilon_k=\sqrt{t^2+\mu^2+2t\mu\cos{k}}$ and $\theta_{k}=\arctan{[t\sin{k}/(t\cos{k}+\mu)]}/2$.  The topological invariant (winding number) that measures the number of edge modes at each end of the wire reads \cite{Chakravarty11}:
\begin{equation}
 P=\oint\frac{d\theta_{k}}{2\pi}
\label{topnumber}
\end{equation} 
and gives $P=0$ $(1)$ for $t<|\mu|$ ($t>|\mu|$), implying the absence (existence) of Majorana end modes for a system with open boundaries. We will now analyze this quantity in the presence of the cavity.

{\it Cavity-coupled wire.}|The combined system Hamiltonian can be mapped to the well-known Dicke model, namely a set of spins interacting with the same photonic mode. We start with the  bulk system, later focusing on a finite wire. In the continuum limit, we can write the interaction Hamiltonian as $H_{\rm int}=\alpha\sum_k\delta_k(a^{\dagger}+a)\tau_k^z$, with $\delta_k=1+(\beta/\alpha)\cos{k}$, which resembles  the interaction Hamiltonian between a photonic mode and a collection of spins: the Dicke model. To analyze the effect of this term on the spectrum,  we consider both the high-frequency ($\omega\gg t+|\mu|$, i.e., much larger than the quasiparticle band width) and the resonant ($|t-|\mu||<\omega<t+|\mu|$) regimes.

\begin{figure}[t]
\begin{center}
\includegraphics[width=0.99\linewidth]{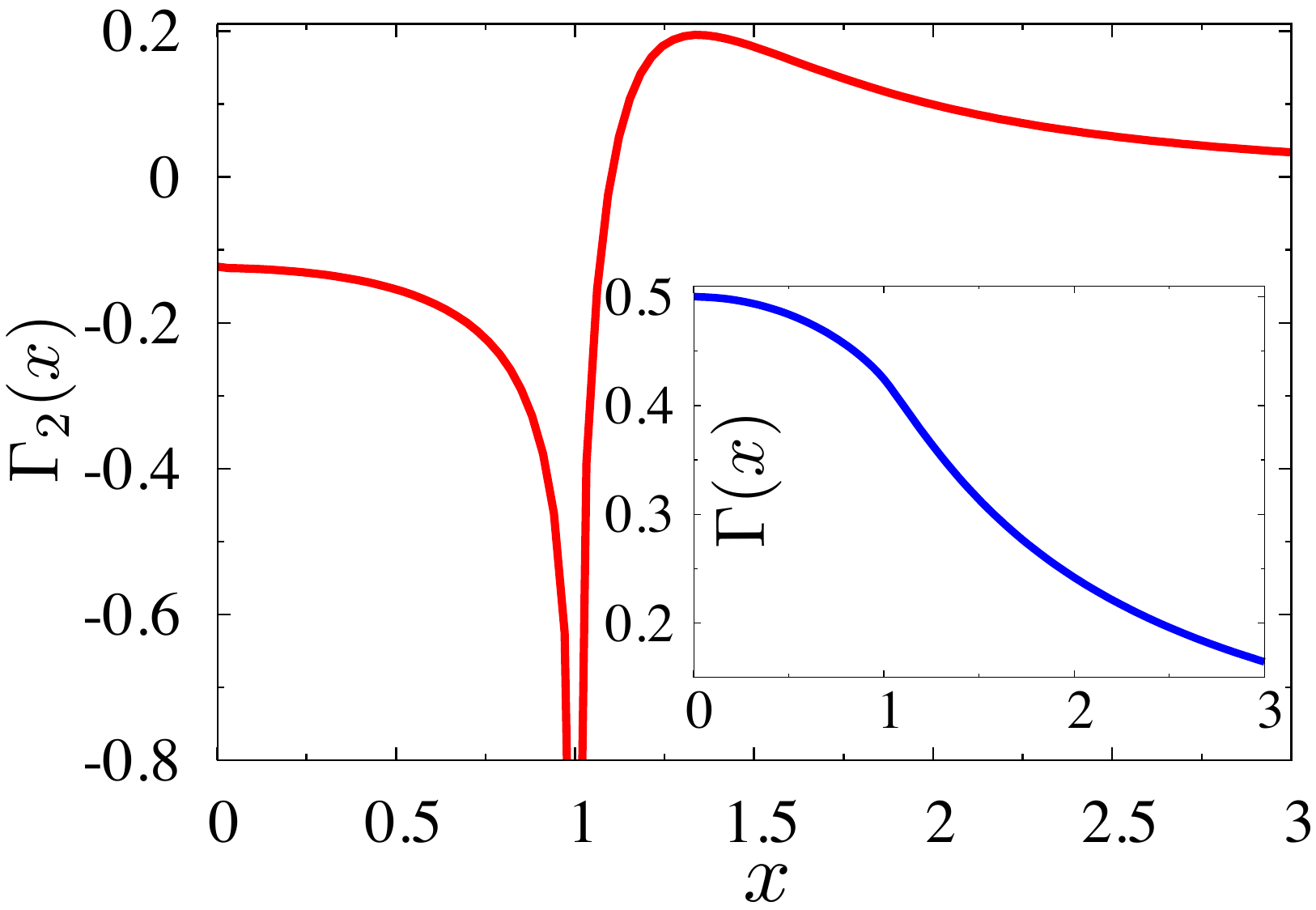}
\caption{The dimensionless functions $\Gamma(x)$ (inset) and $\Gamma_2(x)$ as a function of $x\equiv|\mu|/t$. There is a strong signature of the transition point $x=1$ (i.e., $|\mu|=t$) in $\Gamma_2$, which becomes singular.}
\label{omega_shift_plot}
\end{center}
\end{figure} 

{\it Off-resonant coupling.}|For a large cavity detuning, $\omega\gg t+|\mu|$, all the transitions are off-diagonal, and we can employ second-order perturbation theory using the Schrieffer-Wolff formalism. This means diagonalizing the Hamiltonian in a decoupled basis of the wire-cavity system via a unitary transformation, $\widetilde{\mathcal{H}}=\exp{(S)}\mathcal{H}\exp{(-S)}=\mathcal{H}+[S,\mathcal{H}]+\mathcal{O}(S^2)$, with $S^{\dagger}=-S$ being an anti-Hermitian operator.  We leave the details for the Supplementary Material (SM) and only mention that, although in this situation the photon field cannot easily change the topology of the system, one can utilize the coupling to measure the topological phase of the system\cite{Pedrocchi:2011} as well as the transition point. This is encoded in the cavity frequency shift $\delta\omega=4Nt(\alpha/\omega)^2\,\Gamma(\mu/t)$ with 
\begin{equation}
\Gamma(\mu/t)\simeq\frac{1}{N}\sum_k\delta_k^2\sin k\langle\tau_k^x\rangle\approx\frac{1}{2\pi}\int_{-\pi}^\pi dk\delta_k^2\sin k\langle\tau_k^x\rangle\,,
\label{omega_shift_eq}
\end{equation} 
where $\langle{\dots}\rangle$ means averaging over the (mean-field) ground state of the 1D system. The perturbative treatment is justified for finite-size chains with $\alpha\sqrt{N}\ll\omega$. Moreover, we note that besides the frequency shift, the cavity leads to a shift of the effective chemical potential, which is small in the perturbative regime (see SM).   In Fig.~\ref{omega_shift_plot}, we plot $\Gamma(x)$ and  $\Gamma_2(x)\equiv\partial^2\Gamma/\partial x^2$ as a function of the chemical potential $\mu$. We see that while  $\Gamma(x)$ varies smoothly, $\Gamma_2(x)$ becomes singular at the topological transition point, the divergence  being logarithmic in nature, i.e., $\Gamma_2(1\pm\epsilon)\propto\log{|\epsilon|}$. The corresponding detection method for the transition would thus provide an optical alternative to the more conventional transport-based proposals such as that of Ref.~\cite{Fulga:2011zr}.

In order to experimentally resolve the frequency shift, the condition $\delta\omega/\omega\gg Q^{-1}$ must be met, with $Q$ being the quality factor of the cavity. If we assume microwave cavities with $Q\sim10^6$ \cite{WSB+04}, and taking $t/\omega\sim10^{-1}$ and $\alpha\sqrt{N}/\omega\sim10^{-2}$, we get $\delta\omega/\omega\sim10^{-5}\gg Q^{-1}$. Note that even if the coupling parameter $\alpha$ to the individual bonds turns out to be small, by assuming a large enough chain (i.e., large $N$) one can drastically enhance the effective coupling $\alpha\sqrt{N}$ to reach the strong coupling regime. Moreover, since the cavity is highly detuned from any transitions in the wire, no additional relaxation mechanism are introduced by this coupling.

As an extension of our original model, Eq.~\eqref{Hamiltonian}, we may consider an alternative interaction Hamiltonian, $H_{\rm int}=g\sum_k(\tau_k^+a+a^{\dagger}\tau_k^-)$, where $\tau_k^\pm=(\tau_k^x\pm i\tau_k^y)/2$, which does directly affect the topology of the system. Note that we do not suppose here a rotating-wave approximation (see below). To the second order in the coupling strength $g$, we find: 
\begin{equation}
\Delta H_{\rm int}=\frac{g^2}{\omega}\left[\sum_k\left(a^{\dagger}a+\frac{1}{2}\right)\tau_k^z+\frac{1}{2}\sum_{k\neq k'}(\tau_k^+\tau_{k'}^-+\tau_k^-\tau_{k'}^+)\right]\,.
\label{2nd_order}
\end{equation}
The first term here accounts for a cavity-induced shift of the chemical potential, while the flip-flop coupling in the second term leads  to only a small mean-field chemical-potential shift in the ground state, similarly to the previous section (see SM).  With this effective interaction, it is possible to switch between topologically trivial and nontrivial states by changing the photon number (which is conserved). For example, neglecting the second term in Eq.~(\ref{2nd_order}), the first term leads to a shift in the chemical potential, $\mu\rightarrow\mu_{\rm eff}=\mu-(g^2/\omega)(n_{\rm ph}+1/2)$, with $n_{\rm ph}\equiv\langle a^{\dagger}a\rangle$ (assuming henceforth that $\mu>0$). If the system is in the nontopological phase in the absence of the cavity coupling (i.e., $\mu>t$), for $n_{\rm ph}>n_{\rm ph}^{(c)}=(\mu-t)(\omega/g^2)$ the effective chemical potential satisfies $\mu_{\rm eff}<t$, thus the system crosses over to the topological phase (we neglected the $1/2$ part since typically this transition would happen for $n_{\rm ph}\gg1$). This shows the attractiveness of such  a modified Hamiltonian, which is a quantized analogue of the one proposed in Ref.~\cite{Kitagawa:2011fk}. This type of Hamiltonian can, however, be physically implemented only if the coupling to the cavity leads to changes of the pairing field.

\begin{figure}[t]
\begin{center}
\includegraphics[width=0.99\linewidth]{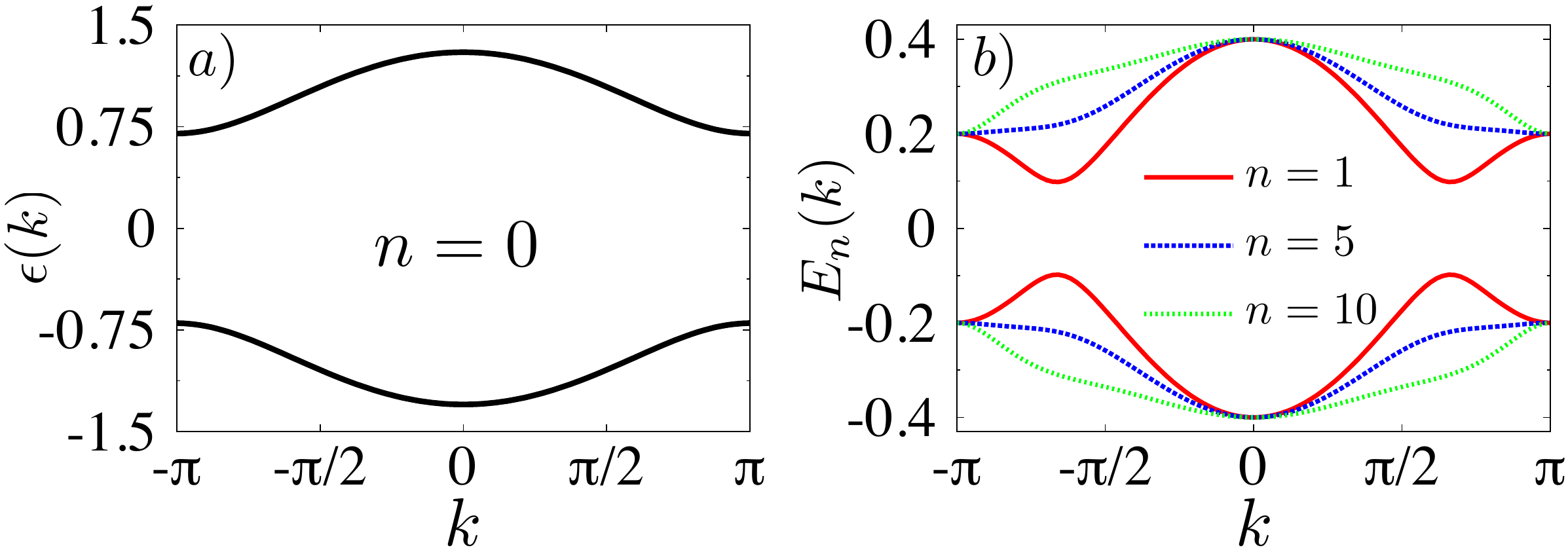}
\caption{{\it Left}:  single-particle spectrum $\epsilon(k)$ for zero photons in the cavity (or in the absence of the cavity). {\it Right}: single-particle spectrum in the presence of the cavity coupling, $E_{n}(k)$, for differerent number $n$ of photons in the cavity. For the plots we used: $\mu=0.3$, $\alpha=0.1$, $\beta=0$, and $\omega=0.9$, all in units of $t$. Note that these parameters put us in the resonant regime.}
\label{SpectrumWithCavity}
\end{center}
\end{figure}

{\it Resonant coupling.}|In the resonant  regime, it is possible to change the system topology at the single-particle level of Eq.~(\ref{Hamiltonian}) depending on the cavity state. To see this, we  first simplify Hamiltonian (\ref{Hamiltonian}) by neglecting the ``counter-rotating" terms that are off-resonant. Leaving the details of the derivation for the SM, we show here only the final result [in the rotated particle-hole basis as described below Eq.~(\ref{BogoGen})]: $\Delta H_{\rm int}=\sum_k\alpha_k^x(\tau_k^+a+a^{\dagger}\tau_k^-)$, where $\alpha_k^x=\alpha\delta_k\sin{(2\theta_k)}$. This closely resembles the original Dicke Hamiltonian \cite{D54}. However, note that both the single-pseudospin splittings as well as the couplings are $k$ dependent (as opposed to the model in the previous section). For each individual $k$, the Hamiltonian reduces to the Jaynes-Cummings Hamiltonian, which is block-diagonal, with each block being a $2\times2$ matrix. Each of these blocks is associated with a given value of the conserved quantity $C_k=\tau_k^z+a^{\dagger}a$, so that the $2\times2$ block acts in the subspace $\{|{\uparrow_k}\rangle\otimes|n\rangle,|{\downarrow_k}\rangle\otimes|n+1\rangle\}$. This  results in an effective two-band Hamiltonian, which we can write in terms of a pseudospin $\bm{s}=(s_x,s_y,s_z)$ (Pauli matrices) as follows:
\begin{eqnarray}
H_{n}(k)&=&\frac{1}{2}(\epsilon_k-\omega)s_z+\frac{\sqrt{n}\alpha_k^x}{2}s_x\,,
\label{effJCHam}
\end{eqnarray}
giving the spectrum $E_{n}(k)=\pm\sqrt{(\omega-\epsilon_k)^2+n(\alpha_k^x)^2}/2$. Note  that  we neglected a constant shift $\Delta E_n=(n+1/2)\omega$ of the two bands for a given $C_k=n$. For each block labelled by $n$, the spectrum has a band gap given by the strength of the pseudospin-photon interaction $\alpha_k^x$ times $\sqrt{n}$. This dramatically alters the initial spectrum $\epsilon_k$, as $n$ increases, which is depicted in Fig.~\ref{SpectrumWithCavity}.

The single-particle picture described above is intimately related to the  recent developments on the topology in periodically-driven systems \cite{Lindner:2011hc,Kitagawa:2010ij,Jiang:2011bs}. This connection is made concrete by considering the classical limit of a cavity prepared in a coherent state with  $n\gg1$. Neglecting the photon-number fluctuation, the single-electron spectrum then becomes $E_\pm(k)=\pm\sqrt{(\omega-\epsilon_k)^2+\Omega_k^2}/2$, with $\Omega_k\equiv\sqrt{\langle n\rangle}\alpha^x_k$ being the classical Rabi frequency and $\langle n\rangle$ the average number of photons. The bulk topology can be correspondingly characterized by Eq.~(\ref{topnumber}), with $\theta_k=\arctan{\left[\Omega_k/(\omega-\epsilon_k)\right]}/2$. We find $P=1$ for $t-|\mu|<\omega<t+|\mu|$, and zero otherwise. Thus the topology of the electron system can be modified due to the interaction with the cavity. 

The preceding results were found within a single-particle picture, which in fact loses its precise meaning in the many-body Dicke-model regime we are actually dealing with. Namely, all the  pseudospins simultaneously interact with the same cavity mode, and the exact many-body spectrum involves all of them. In order to simplify further progress, we notice that the interacting Hamiltonian admits a conserved global quantity $C=\sum_{k}\tau_k^z+a^{\dagger}a$. The simplest situation is when  there are no photons in the cavity and $C=-N/2$ reaches its lowest possible value. The combined eigenstate of the system is just the product state $|\psi_{\rm tot}^0\rangle=|{\Downarrow}\rangle\otimes|0\rangle$ where $|{\Downarrow}\rangle\equiv|{\downarrow}\dots\downarrow\rangle$, and the energy $E_0=-\sum_{k}\epsilon_k$. For a finite number of photons in the cavity the problem becomes progressively more involved, and reasonably simple results can only be found for $n=0,1,2$, as well as for $n\gg1$ (mean field), with $n=C+N/2$.

We next wish to analyze the many-body ground state (at a fixed-$n$ subspace) when the cavity is populated with photons. We focus only on the   $n=1$ case, as this already contains much of the relevant physics. In this case, the wave function can be expanded as
\begin{equation}
|\psi_{\rm tot}^1\rangle=a|{\Downarrow}\rangle\otimes|1\rangle+\sum_{k}b_k|{\Downarrow;\uparrow_k}\rangle\otimes|0\rangle\,,
\label{EntWF}
\end{equation}
where the coefficients $a$ and $b_k$'s are found by solving the Schrodinger equation $\mathcal{H}|\psi_{\rm tot}^1\rangle=E_1|\psi_{\rm tot}^1\rangle$\cite{TsyplyatyevLoss:2010}.
The simplest case is  when $\mu=0$, so that $\epsilon_k\equiv t$, and  the ground-state energy is found to be  $E_1=-(N-1)t+\omega/2-\sqrt{N\Gamma_1+(t-\omega/2)^2}$, with $\Gamma_1=\sum_k(\alpha_k^x)^2/N$. (For a finite $\mu\neq0$, the result is more complicated, albeit similar in nature.) This means the ground-state energy is lowered at resonance compared to the non-interacting situation by the amount $\propto\sqrt{N}$. If the number of pseudospins $N$ (or the strength of coupling $\Gamma_1$) exceeds a critical value, the absolute ground state thus becomes populated with photons. For example, in the homogeneous coupling described  above $E_1=E_0$ for $\Gamma_1=\omega^2/N$ at resonance. Although this transition is not innately topological, it can switch the topology of the system. Thus, at the critical coupling, the system switches to a different ground state populated by photons, and the topology, which is related to the appearance of end modes, must be appropriately reexamined. To that end, however, we need to calculate the total many-body energy in a finite system, which is beyond the single-particle bulk reasoning.

{\it Finite system.}|The Majorana states are actually boundary modes, even though their existence can be inferred from the bulk spectrum. We use the discrete lattice model to numerically diagonalize the Hamiltonian in the zero- and one-photon regimes, in order to explicitly identify zero modes. We have previously noted the existence of Majorana end modes of a free chain for $\mu=0$. That result is extended to $\mu\neq0$, after finding an orthogonal transformation $\mathcal{M}$ such that $\widetilde{H}_{1D}=\mathcal{M}H_{1D}\mathcal{M}^T=i\sum_{m=1}^N\epsilon_m\widetilde{\gamma}_m^1\widetilde{\gamma}_m^2$, with $\widetilde{\gamma}_{m}^{p}=\sum_js_{mj}^{pr}\gamma_j^r$ (here, $j,m=1,\dots,N$; $p,r=1,2$), $s_{mj}^{pr}(t,\mu)$ being the elements of the real orthogonal $2N\times2N$ matrix $\mathcal{M}$ \cite{Kitaev:2001qf}. We can write also $\widetilde{H}_{\rm 1D}=\sum_{m=1}^N\epsilon_m(\widetilde{c}^{\dagger}_m\widetilde{c}_m-1/2)$, with $\widetilde{c}^{\dagger}_m=(\widetilde{\gamma}^1_m+i\widetilde{\gamma}^2_m)/2$, and the eigenenergies $\epsilon_n$ define the spectrum of the chain which contain the zero modes for $|\mu|<t$. 

Focusing on the topological regime, we now analyze the evolution of the many-body zero mode as the first photon is starting to populate the cavity.  We assume the cavity frequency to satisfy $(t-|\mu|)<\omega<2(t-|\mu|)$, which allows for the resonant transitions only from the Majorana states to the gapped fermionic continuum. Note that the parity of the system, $p\equiv\gamma_{1}^1\dots\gamma_N^2$, is conserved, i.e., $[p,\mathcal{H}]=0$,  which means this is a good quantum number even in the presence of the cavity. The cavity may thus affect the splitting between the parities, but not mix different parities. The interaction Hamiltonian can be written in terms of  the operators $\widetilde{c}^{\dagger}_m$($\widetilde{c}_m$) instead of $\tau_k$'s which, in the rotating-wave approximation, becomes $H_{\rm int}=\sum_{m\neq m_F}A_mf(\widetilde{c}_F,\widetilde{c}_F^{\dagger})\widetilde{c}^{\dagger}_ma+{\rm H.c.}$, with $A_m$ the effective coupling and  $f(\widetilde{c}_F,\widetilde{c}_F^{\dagger})$ being a linear function in $\widetilde{c}_F,\widetilde{c}_F^{\dagger}$. As in the pseudospin model, here too we can use the number of excitations $C\equiv\sum_{n}\widetilde{c}_m^{\dagger}\widetilde{c}_m+a^{\dagger}a$ as a conserved quantity for each parity $p$ to find the spectrum and, as an example, we calculate the energy of the system for $C=1$. The corresponding eigenvalues for the two parities are found from solving the eigenvalue problem similarly to Eq.~(\ref{EntWF}), which boils down (see SM for details) to solving the equation
\begin{equation}
\left(\omega-\epsilon_{+(-)}-\epsilon_{\rm tot}^0\right)+\sum_{m,r,r',p,p'}\frac{(-1)^{r+r'(p')}S_m^{rr'}S_m^{pp'}}{\epsilon_{+(-)}+\epsilon_{\rm tot}^0-2\epsilon_{m}}=0\,,
\label{OnePhotEqn}
\end{equation}
where $\epsilon_{\pm}$ stand for the energy of the states of parity $\pm$; $\epsilon_{\rm tot}^0=\sum_j\epsilon_m$; $r,r',p,p'=1,2$; and  $S_{m}^{rr'}=\alpha i^{r+r'}\sum_{j}(s_{jm}^{1r}s_{j0}^{2r'}-s_{j0}^{1r'}s_{jm}^{2r})$ setting $\beta=0$. The splitting between the lowest-energy states of different parities in the subspace $C=1$ is denoted by $\Delta E_{\rm maj}=|\epsilon_{+}^{\rm gs}-\epsilon_{-}^{\rm gs}|$, while the gap to the continuum by $\Delta E_{\rm gap}=\min{(\epsilon_{\pm}^{\rm es})}-\max{(\epsilon_{\pm}^{\rm gs}})$, with $\epsilon_{\pm}^{\rm gs(es)}$ being the ground-state (excited-state) energies. Note that such a definition can be generalized for any $C\geq 1$, but not for $C=0$.   

\begin{figure}[t]
\begin{center}
\includegraphics[width=1\linewidth]{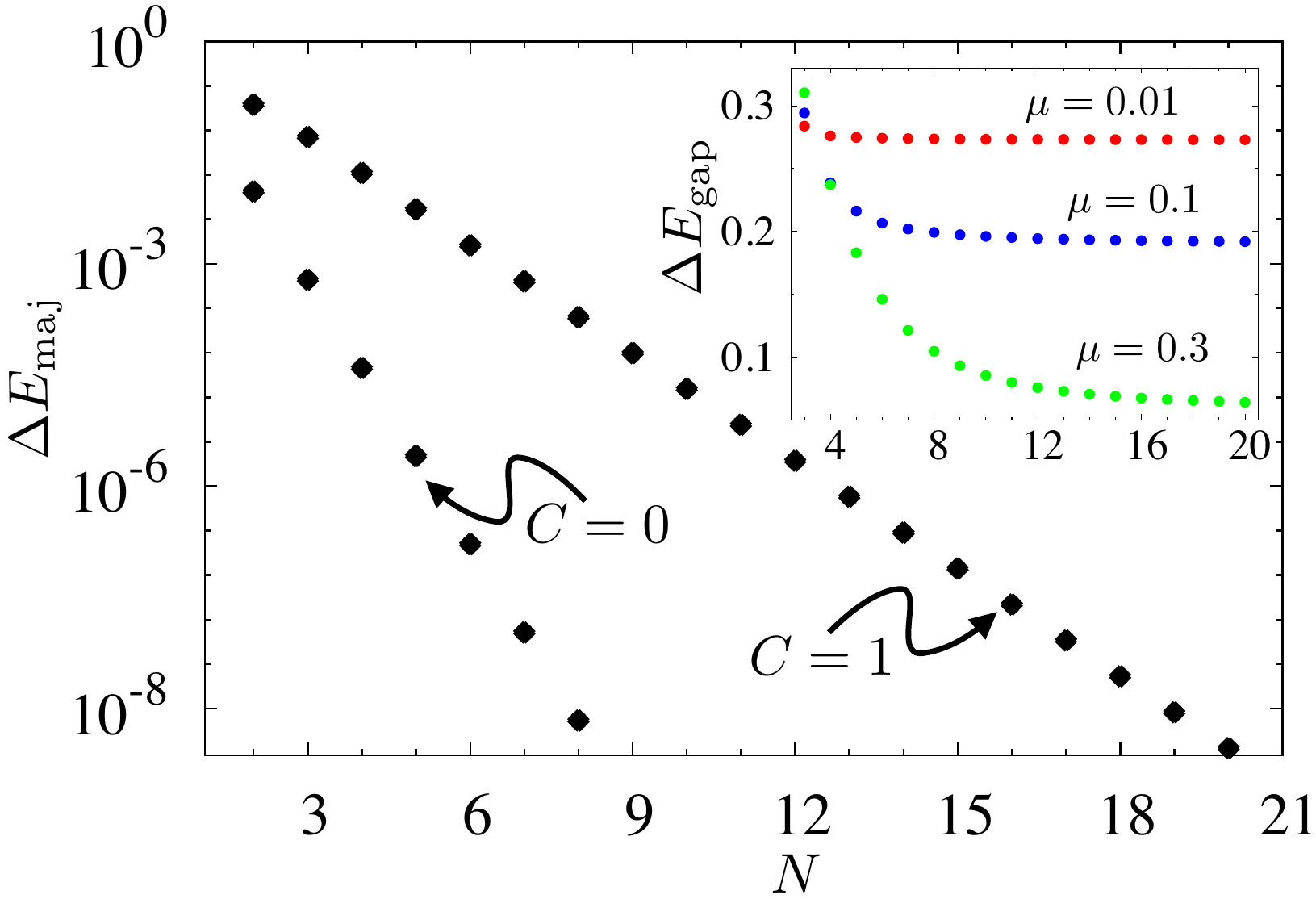}
\caption{{\it Main}: The energy splitting of the two lowest-energy states, $\Delta E_{\rm maj}$, for zero ($C=0$) and one excitation ($C=1$), as a function of the number of sites $N$. For these plots, we used $\mu=0.1$, $\alpha=0.1$, $\beta=0$, and $\omega=1$, all in units of $t$. {\it Inset:} The effective gap from the Majorana states to the continuum, $\Delta E_{\rm gap}$, for $C=1$, as a function of $N$ (using here $\alpha=0.2$ to exaggerate trends).}
\label{MajSplit}
\end{center}
\end{figure} 

Figure~\ref{MajSplit} shows the energy splitting between the lowest energy states, $\Delta E_{\rm maj}$, for $C=0$ and  $C=1$, as well as the splitting (gap) between these two states and the continuum, $\Delta E_{\rm gap}$ for $C=1$. We see that in both cases the splitting between the two states scales exponentially with the number of sites, although slower for $C=1$. Interestingly, the energy gap $\Delta E_{\rm gap}$ saturates at a value $\propto\alpha$ for large $N$.
Physically, the resonant interaction  brings the Majorana modes into the continuum spanned by the $C=1$ states, but as this interaction increases, precisely one state per parity is pushed below the continuum  and span a well-defined two-dimensional degenerate subspace. These degenerate modes are highly entangled electron-photon states, which we call Majorana polaritons. Note that there is no Dicke-like enhancement $\sqrt{N}\alpha$ of $\Delta E_{\rm gap}$ for large $N$ (see $E_1$ below Eq. (\ref{EntWF})), since it is exactly canceled by the overlap of the Majorana states with the continuum, which scales as $1/\sqrt{N}$. Nevertheless, for large enough $\alpha$, the Majorana polaritons are well defined.

{\it Summary.}|To conclude, we analyzed the spectrum of a 1D Kitaev chain coupled to a microwave cavity in both off- and on-resonant regimes. An off-resonant cavity can serve as a measuring device for the topological transitions, while in  the resonant regime the topology can be changed depending on the photon number. We show that at the (mean-field) single-particle level the spectrum maps to that  of classically-driven (Floquet) topological insulators. In the many-body picture, we find that the spectrum can exhibit a degenerate ground state even for a finite number of photons (despite strong hybridization of the original Majoranas with the continuum). These states are entangled electron-photon composites interpreted as Majorana polaritons. Our model can be relevant for Ising spin chains \cite{TserkovnyakLoss:2011}, one-dimensional nanowires with strong SOI in the presence of an $s$-wave superconductor \cite{Sau:2010vn}, as well as in the context of cold-atom physics \cite{Jiang:2011bs}.  

This work was supported by the NSF under Grant No. DMR-0840965, the Alfred P. Sloan Foundation, and by DARPA. We gratefully acknowledge fruitful discussions with Daniel Loss and Manuel Schmidt.

\bibliography{MajoranaCavityArchive}

\begin{widetext}

\section*{Chain-photon coupling in second-order perturbation theory}     

In this Supplementary Material we give a detailed description of the second-order calculation for the two types of the wire-cavity coupling Hamiltonians. We will focus on the coupling via the chemical potential since the other type of coupling (via the modification of $t$) turns out to be a straightforward extension. The BdG+cavity Hamiltonian in the $k$ space reads $\mathcal{H}=H_{\rm BdG}+H_{\rm int}+H_{\rm ph}$:
\begin{equation}
H_{\rm BdG}=\sum_k\bm{n}_k\cdot\bm{\tau}_k\,,\,\,\,H_{\rm int}=\sum_k\alpha_k\tau_k^z(a^{\dagger}+a)\,,\,\,\,H_{\rm ph}=\omega\,a^{\dagger}a\,,
\end{equation}
where $\bm{n}_k=\left(-t\sin{k}, 0, -(t\cos{k}+\mu)\right)$ and $\alpha_k=\alpha+\beta\cos{k}$. $t>0$ is the hopping parameter, $\mu$ is the chemical potential,  $\omega$ is the intrinsic cavity frequency, and $\alpha$ ($\beta$) are is wire-cavity coupling strengths via modification of the chemical potential (hopping parameter). We consider the off-resonant regime, $\omega\gg t+|\mu|$, and assume the frequency to be large enough for the interaction term to be treated perturbatively. The precise condition for the applicability of the perturbation theory will be discussed below. We use the Schrieffer-Wolff transformation to derive the second-order corrections. This implies finding an anti-Hermitian operator $S$ such that
\begin{equation}
\widetilde{\mathcal{H}}=e^{S}\mathcal{H}e^{-S}=\mathcal{H}+[S,\mathcal{H}]+\frac{1}{2}[S,[S,\mathcal{H}]]+\dots\,,
\end{equation}
is diagonal in the eigenbasis of $H_{\rm BdG}+H_{\rm ph}$. For the linear in $\alpha_k$ terms to vanish, we require $H_{\rm int}=[H_{\rm BdG}+H_{\rm ph},S]$ or
\begin{equation}
S=\mathcal{L}^{-1}H_{\rm int}\equiv i\lim_{\epsilon\rightarrow0}\int_0^{\infty}ds e^{-\epsilon s}e^{-i (H_{\rm BdG}+H_{\rm ph})s}H_{\rm int}e^{i(H_{\rm BdG}+H_{\rm ph})s}\,,
\end{equation}
where the (super)operator $\mathcal{L}$ is defined formally by $\mathcal{L}\,A=[H_{\rm BdG}+H_{\rm ph},A]$, for an arbitrary operator $A$. This gives:
\begin{align}
S&=i\lim_{\epsilon\rightarrow0}\sum_k\int_0^{\infty}ds \alpha_k\underbrace{e^{-i\bm{n}_k\cdot\bm{\tau}_ks}\tau_k^ze^{i\bm{n}_k\cdot\bm{\tau}_ks}}_{T_k(s)}(a^{\dagger}e^{i\omega s}+ae^{-i\omega s})\nonumber\\
&\approx i\sum_k\frac{\alpha_k}{\omega}\left[i\tau_k^z(a^{\dagger}-a)-\frac{2t\sin{k}}{\omega}\tau_k^y(a^{\dagger}+a)\right]\,,
\end{align}
disregarding terms of order $\omega^{-3}$. This is seen by explicitly evaluating $T_k(s)$:
\begin{equation}
T_k(s)=\tau_k^z\cos{(2\epsilon_ks)}+\frac{t\sin{k}}{\epsilon_k}\sin{(2\epsilon_ks)}\tau_k^y+2\frac{t\cos{k}+\mu}{\epsilon_k}\left(\frac{t\sin{k}}{\epsilon_k}\tau_k^x+\frac{t\cos{k}+\mu}{\epsilon_k}\tau_k^z\right)\sin^2{(\epsilon_ks)}\,,
\end{equation}
where $\epsilon_k\equiv|\bm{n}_k|$. Inserting this $T_k(s)$ in the integral over time leads to the following result for $S$:
\begin{align}
S&=i\sum_k\frac{\omega\alpha_k}{\omega^2-4\epsilon_k^2}\left[i\tau_k^z(a^{\dagger}-a)-\frac{2t\sin{k}}{\omega}\tau_k^y(a^{\dagger}+a)-i\frac{4(t\cos{k}+\mu)}{\omega^2}\bm{n}_k\cdot\bm{\tau}_k(a^{\dagger}-a)\right]\nonumber\\
&\approx-\sum_k\frac{\alpha_k}{\omega}\left[\tau_k^z(a^{\dagger}-a)+i\frac{2t\sin{k}}{\omega}\tau_k^y(a^{\dagger}+a)\right]
\end{align}
for $\omega\gg t+|\mu|$, which allows us to neglect the $1/\omega^3$ terms. Now we are in position to derive the effective Hamiltonian within second order, $\Delta H_{\rm int}=[S,H_{\rm int}]/2$, which is given by:
\begin{align}
\Delta H_{\rm int}&=-\frac{1}{2}\sum_{k,k'}\frac{\alpha_k\alpha_{k'}}{\omega}\left([\tau_{k'}^z(a^{\dagger}-a),\tau_k^z(a^{\dagger}+a)]+i\frac{2t\sin{k}}{\omega}[\tau_{k'}^y(a^{\dagger}+a),\tau_k^z(a^{\dagger}+a)]\right)\nonumber\\
&=\sum_{k,k'}\frac{\alpha_k\alpha_{k'}}{\omega}\tau_k^z\tau_{k'}^z+\sum_k\frac{4t\alpha_k^2\sin{k}}{\omega^2}\left(a^{\dagger}a+\frac{1}{2}\right)\tau_{k}^x\,.
\label{eff_ham}
\end{align}
This is the Hamiltonian used in the Main Text (MT). We identify two distinct contributions: the purely electronic first term that introduces many-body effects and the second term (dependent on the photon number) that renormalizes the $p$-wave pairing strength. Note that the off-diagonal terms $\propto\alpha_k^2[(a^\dagger)^2+a^2]$ were dropped here, since their subsequent Schrieffer-Wolff elimination would produce higher-order terms that are outside of our scope.

\section*{Mean-field solution for frequency shift}

In the MT, we argued that the second term in the Hamiltonian, Eq.~(\ref{eff_ham}), leads to a shift in the cavity frequency that can be measured, allowing to identify the transition point as well as the phase the 1D Kitaev chain is in. The corresponding expression for the frequency shift is
\begin{equation}
\delta\omega=\frac{4t}{\omega^2}\sum_k\alpha_k^2\sin{k}\,\tau_k^x\,.
\end{equation} 
Performing then the substitution $\tau_k^x\rightarrow\langle\tau_k^x\rangle$, which neglects electronic fluctuations, we obtain a frequency shift $\langle\delta\omega\rangle$ (that is independent of the photon number if neglect the small backaction of this coupling on the electronic system).  However, we did not explicitly evaluate the average $\langle\tau^x_k\rangle$, and, moreover, did not discuss how the first term in Eq.~(\ref{eff_ham}) was accounted for. Note that the first term is $\propto1/\omega$, while the second $\propto1/\omega^2$, i.e., much smaller in the high-frequency regime of interest. The first term,
\begin{align}
\Delta  H_{\rm int}'&=\sum_{k,k'}\frac{\alpha_k\alpha_{k'}}{\omega}\tau_k^z\tau_{k'}^z\,,
\end{align}
is also special in that it leads to interaction effects. Let us first consider it separately. In a mean-field approximation, the many-body effects are eliminated by the substitution:
\begin{equation}
\tau_k^z\tau_{k'}^z\simeq\langle\tau_k^z\rangle\tau_{k'}^z+\tau_k^z\langle\tau_{k'}^z\rangle-\langle\tau_k^z\rangle\langle\tau_{k'}^z\rangle\,.
\end{equation}
We assume, furthermore, that $\alpha_k\equiv\alpha$ (i.e., there are no tunneling effects). This mean-field interaction can thus be accounted for by a chemical potential shift, namely $\mu\rightarrow\mu_{\rm eff}=\mu-A$ with
\begin{equation}
A=\frac{2\alpha^2}{\omega}\sum_{k}\langle\tau_k^z\rangle\,,
\end{equation}
where the average $\langle{\dots}\rangle$ is taken self-consistently over the renormalized states. Neglecting the frequency-shift term $\propto1/\omega^2$ in Eq.~(\ref{eff_ham}) (which could also be treated exactly if necessary), we can write the renormalized BdG Hamiltonian as follows:
\begin{equation}
H_{\rm BdG}^{\rm eff}=-\sum_k\left[(t\cos{k}+\mu_{\rm eff})\tau_k^z+t\sin{k}\tau_k^x\right]\,,
\end{equation}
whose eigenenergies are $\epsilon_k^{\rm eff}=\sqrt{t^2+\mu_{\rm eff}^2+2t\mu_{\rm eff}\cos{k}}$. The expectation values for $\tau_{k}^{x,y}$ in the ground state of this Hamiltonian read:
\begin{equation}
\langle\tau_k^x\rangle=\frac{t\sin{k}}{\epsilon_k^{\rm eff}}\,,\,\,\,\langle\tau_k^z\rangle=\frac{t\cos{k}+\mu_{\rm eff}}{\epsilon_k^{\rm eff}}\,,
\end{equation}
which, in turn, enter in the self-consistent equation for $A$:
\begin{equation}
0=A-\frac{2\alpha^2}{\omega}\sum_k\frac{t\cos{k}+\mu-A}{\epsilon_k^{\rm eff}}\to A-\frac{\alpha^2N}{\pi\omega}\int_{-\pi}^{\pi} dk\frac{t\cos{k}+\mu-A}{\epsilon_k^{\rm eff}}\,.
\end{equation}
Here, we transformed the sum into an integral, supposing a large number of sites, $N\gg1$. In Fig.~\ref{implicit}, we plot this $A$ as a function of the chemical potential for several values of the dimensionless wire-cavity coupling parameter $g\equiv\alpha^2N/\omega t$. We see that $A$ follows a linear dependence for $|\mu|\ll t$ and it saturates for $|\mu|\gg t$ to values that are different depending on the coupling parameter $g$. Actually, we can obtain analytical results for $A$ in the two limiting cases.  For small $|\mu|\ll t$, the asymptotic form is $A=r\mu$ with the slope:
\begin{equation}
r=\frac{g}{1+g}~~~\to~~~\mu_{\rm eff}=(1-r)\mu=\frac{\mu}{1+g}\,,
\end{equation}
which means the function $A$ can be as large as $\mu$ for large $N$, but not exceed it. In this case, the cavity reduces the chemical potential keeping the system in the topological phase, i.e., it stabilizes it by shifting the transition point. In the opposite (nontopological) case, $|\mu|\gg t$, the function $r$ becomes instead:
\begin{equation}
r=\frac{g}{\mu}\left(2-\frac{t^2}{2\mu^2}\right)~~~\to~~~\mu_{\rm eff}=\mu-g\left(2-\frac{t^2}{2\mu^2}\right)\,.
\end{equation} 

\begin{figure}[t]
\begin{center}
\includegraphics[width=0.5\linewidth]{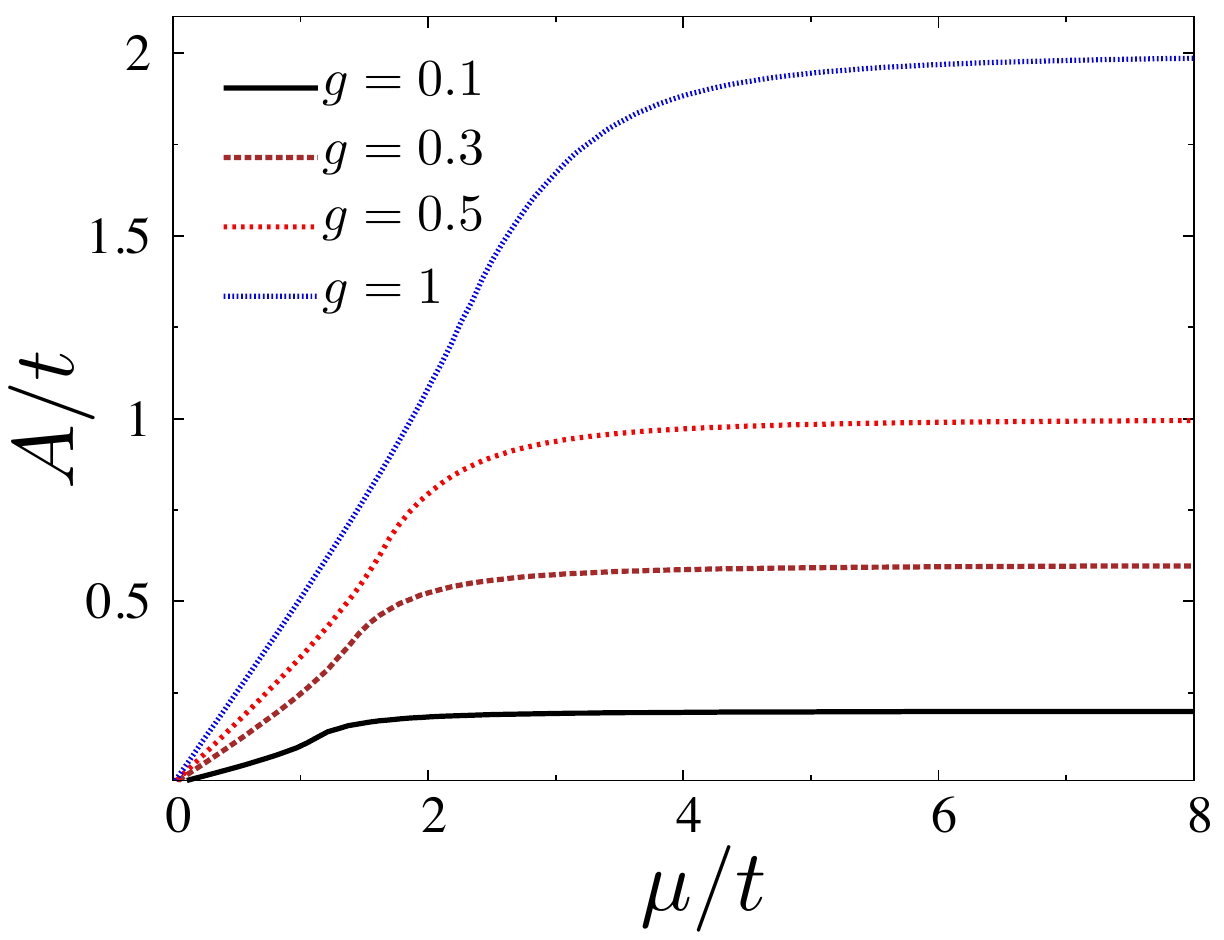}
\caption{The cavity-induced  chemical potential shift $A$ as a function of the bare chemical potential $\mu$ (in units of $t$) for different values of the effective coupling strength $g=\alpha^2N/t\omega$.}
\label{implicit}
\end{center}
\end{figure}

For the perturbation theory to be internally consistent, we need to assume $\omega\gg\alpha\sqrt{N}$, which makes the second term in Eq.~(\ref{eff_ham}) small compared to the hopping in $H_{\rm BdG}$. Concerning the first term in Eq.~(\ref{eff_ham}), it is of order $(\alpha\sqrt{N}/\omega)^2\omega\ll\omega$ if $\alpha\sqrt{N}\ll\omega$, which can, nevertheless, be still comparable or even larger than $t$ and/or $|\mu|$. A more stringent condition, $g\ll1$, would assure that $\Delta H_{\rm int}$ causes only small shifts in the electronic spectral properties. We can finally express the frequency shift $\delta\omega$ using the self-consistent wave functions to find:
\begin{equation}
\langle\delta\omega\rangle=\frac{4t^2\alpha^2}{\omega^2}\sum_k\frac{\sin^2{k}}{\sqrt{t^2+\mu_{\rm eff}^2+2t\mu_{\rm eff}\cos{k}}}=\frac{2t^2g}{\pi\omega}\int_{-\pi}^{\pi}dk\frac{t\sin^2{k}}{\sqrt{t^2+\mu_{\rm eff}^2+2t\mu_{\rm eff}\cos{k}}}\,,
\end{equation}
which reduces to the expression used in the MT. As  can be seen from Fig.~\ref{implicit}, for $g\ll 1$, we can approximate $\mu_{\rm eff}\approx\mu$. In the MT, we wrote $\langle\delta\omega\rangle=4Nt(\alpha/\omega)^2\Gamma(\mu/t)$ and plotted both $\Gamma(x)$ and $\Gamma_2(x)\equiv\partial^2\Gamma/\partial x^2$. While the general  expression for $\Gamma(x)$ is too lengthy to be showed here, there are limiting cases when this becomes very simple. We will thus provide the expressions for $\Gamma(x)$ and $\Gamma_2(x)$ for the regimes $x\ll1$, $x\simeq 1$, and  $x\gg1$. For $\Gamma(x)$, we obtain:
\begin{equation}
\Gamma(x)=\left\{
\begin{array}{cl}
\displaystyle{\frac{1}{2}\left(1-\frac{x^2}{8}\right)}\,, & {\rm for }~~~x\ll1\\
\displaystyle{\frac{2}{3\pi}\left(3-x\right)}\,, & {\rm for }~~~x\simeq 1\\
\displaystyle{\frac{1}{2x}}\,, & {\rm for }~~~ x\gg1
\end{array}
\right.\,,
\end{equation}
which implies that $\langle\delta\omega\rangle_{\mu=0}=2tN(\alpha/\omega)^2$ (maximum), and it scales as $\langle\delta\omega\rangle\sim 1/\mu$ for $\mu\gg t$. This means that $\langle\delta\omega\rangle$ is just a fraction of the hopping parameter $t$. Note also that the function $\Gamma(x)$ is continuos at $x=1$, so there is no jump in the frequency shift. On the other hand, the limiting expressions for $\Gamma_2(x)$ read:
\begin{equation}
\Gamma_2(x)=\left\{
\begin{array}{cl}
\displaystyle{-\frac{1}{8}\left(1+\frac{3}{4}x^2\right)}\,, & {\rm for }~~~x\ll1\\
\displaystyle{\frac{1}{\pi}\log{|x-1|}}\,, & {\rm for }~~~x\simeq 1\\
\displaystyle{\frac{1}{x^3}}\,, & {\rm for }~~~x\gg1
\end{array}
\right.\,,
\end{equation}
which diverges logarithmicaly  as $x\rightarrow1$.

\section*{Rotating-wave approximation for finite-chain spectrum}

In this section, we derive the rotating-wave approximation (RWA) Hamiltonian for a finite chain, as well as the resulting eigenenergy equation for $C=1$ [Eq. (8) in the MT]. We first write the wire and interaction Hamiltonians in terms of Majorana operators:
\begin{align}
H_{\rm 1D}&=-it\sum_{j=1}^{N-1}\gamma_{j}^2\gamma_{j+1}^1-i\mu\sum_{j=1}^N\gamma_{j}^1\gamma_j^2\,.\nonumber\\
H_{\rm int}&=i\left[\beta\sum_{j=1}^{N-1}(\gamma_{j+1}^1+i\gamma_{j+1}^2)(\gamma_j^1+\gamma_{j}^2)+i\alpha\sum_{j=1}^N\gamma_{j}^1\gamma_j^2\right](a^{\dagger}+a)+{\rm H.c}.\,.
\end{align}
We can diagonalize the Hamiltonian in the absence of the cavity using the prescription in Ref.~[1] of the MT and write
\begin{equation}
H_{\rm 1D}=i\sum_{m=1}^{N}\epsilon_m\widetilde{\gamma}_m^1\widetilde{\gamma}_m^2\equiv\sum_{m=1}^N\epsilon_m\left(\widetilde{c}^{\dagger}_m\widetilde{c}_m-\frac{1}{2}\right)\,,
\end{equation}
with $\epsilon_m\geq0$ being the spectrum and 
\begin{equation}
\left(
\begin{array}{c}
\widetilde{\gamma}_{1}^1\\
\widetilde{\gamma}_1^2\\
\vdots\\
\widetilde{\gamma}_N^1\\
\widetilde{\gamma}_N^2
\end{array}
\right)
=M\left(
\begin{array}{c}
\gamma_1^1\\
\gamma_1^2\\
\vdots\\
\gamma_N^1\\
\gamma_N^2
\end{array}
\right)\,,
\end{equation}
in terms of an orthogonal $2N\times2N$ matrix $M$ ($MM^{T}=M^{T}M=I_{2N}$). The fermionic operators are given by $\widetilde{c}^{\dagger}_m=(\widetilde{\gamma}_m^1-i\widetilde{\gamma}_m^2)/2$ and $\widetilde{c}_m=(\widetilde{\gamma}_m^1+i\widetilde{\gamma}_m^2)/2$. 
More specifically, we can write
\begin{equation}
\widetilde{\gamma}_{m}^{p}=\sum_{j=1}^N\sum_{p'=1,2}r_{j,m}^{p,p'}\gamma_{j}^{p'}\,,
\end{equation}
with the coefficients $r_{j,m}^{p,p'}$ being the entries of the matrix $M$ that can be found numerically. Note that if $\epsilon_m$ is a solution, that $-\epsilon_m$ is a solution too. Due to this symmetry, there are possible zero-energy solutions which, if existing, satisfy
\begin{equation} 
r_{j,0}^{1,2}=C^{1,2}\left(-\frac{\mu}{t}\right)^{\pm j}\,,
\end{equation}
with the coefficients $C^{1,2}$ determined by the normalization conditions (not shown, see Ref.~\cite{Kitaev:2001qf}). 
Next, we write
\begin{equation}
\gamma_j^{1,2}=\sum_{m=1}^N\sum_{p=1,2}s_{j,m}^{p,p'}\widetilde{\gamma}_j^{p}\,,
\end{equation}
where $s_{j,m}^{p,p'}\equiv r_{m,j}^{p',p}$ (since $M$ is orthogonal). The spin-photon interaction Hamiltonian is then obtained as follows:
\begin{align}
H_{\rm int}&=\sum_{m,m'}\sum_{p,p'}\left[\beta\sum_{j=1}^{N-1}(r_{j+1,m}^{1,p}+ir_{j+1,m}^{2,p})(r_{j,m'}^{1,p'}-ir_{j,m'}^{2,p'})\right.+\left.i\alpha\sum_{j=1}^Nr_{j,m}^{1,p}r_{j,m'}^{2,p'}\right]\widetilde{\gamma}_{m}^p\widetilde{\gamma}_{m'}^{p'}(a^{\dagger}+a)+{\rm H.c.}\nonumber\\
&=\sum_{m,m'}\sum_{p,p'}A_{m,m'}^{p,p'}[\widetilde{c}^{\dagger}_m+(-1)^{p-1}\widetilde{c}_{m}][\widetilde{c}^{\dagger}_{m'}+(-1)^{p'-1}\widetilde{c}_{m'}](a^{\dagger}+a)+{\rm H.c.}\,,\nonumber\\
A_{m,m'}^{p,p'}&=i^{p+p'-2}\left[\beta\sum_{j=1}^{N-1}(r_{j+1,m}^{1,p}+ir_{j+1,m}^{2,p})(r_{j,m'}^{1,p'}-ir_{j,m'}^{2,p'})\right.+\left.i\alpha\sum_{j=1}^Nr_{j,m}^{1,p}r_{j,m'}^{2,p'}\right]\,.
\end{align}
Next we use the RWA to deal with the resonant regime thus excluding the fast-rotating terms. The coupling Hamiltonian in the interaction picture reads:
\begin{equation}
H_{\rm int}(t)=\sum_{m,m'}\sum_{p,p'}A_{m,m'}^{p,p'}[\widetilde{c}^{\dagger}_me^{i\epsilon_mt}+(-1)^{p-1}\widetilde{c}_{m}e^{-i\epsilon_mt}][\widetilde{c}^{\dagger}_{m'}e^{i\epsilon_{m'}t}+(-1)^{p'-1}\widetilde{c}_{m'}e^{-i\epsilon_{m'}t}](a^{\dagger}e^{i\omega t}+ae^{-i\omega t})\,.
\end{equation}
When the frequency is such that it couples resonantly only the Majoranas to the continuum, but not the two bulk bands, the physics can be described in terms of a Majorana polariton, as outlined in the MT. Thus, we simplify the discussion by performing the RWA to the above Hamiltonian to get:
\begin{align}
H_{\rm int}^{\rm RWA}=&\sum_{m}\sum_{p,p'}\left[A_{0,m}^{p,p'}\left(\widetilde{c}^{\dagger}_0+(-1)^{p-1}\widetilde{c}_{0}\right)\left(\widetilde{c}^{\dagger}_{m}a+(-1)^{p'-1}\widetilde{c}_{m}a^{\dagger}\right)\right.\nonumber\\
&+\left.A_{m,0}^{p,p'}\left(\widetilde{c}^{\dagger}_ma+(-1)^{p-1}\widetilde{c}_{m}a^{\dagger}\right)\left(\widetilde{c}^{\dagger}_{0}+(-1)^{p'-1}\widetilde{c}_{0}\right)\right]\nonumber\\
=&\sum_{m}\sum_{p,p'}\left(A_{0,m}^{p,p'}-A_{m,0}^{p',p}\right)\left(\widetilde{c}^{\dagger}_0+(-1)^{p-1}\widetilde{c}_{0}\right)\left(\widetilde{c}^{\dagger}_{m}a+(-1)^{p'-1}\widetilde{c}_{m}a^{\dagger}\right)\nonumber\\
=&\sum_{m}A_m(\widetilde{c}^{\dagger}_0,\widetilde{c}_{0})\widetilde{c}^{\dagger}_{m}a+{\rm h.c.}\,,\nonumber\\
A_m(\widetilde{c}^{\dagger}_0,\widetilde{c}_{0})=&\sum_{p,p'}\underbrace{\left(A_{0,m}^{p,p'}-A_{m,0}^{p',p}\right)}_{\displaystyle{S_m^{pp'}}}\left(\widetilde{c}^{\dagger}_0+(-1)^{p-1}\widetilde{c}_{0}\right)\,.
\end{align}

Note that  we labeled the zero modes by $m=0$ (for convenience and not meaning the first site). As mentioned in the MT, this Hamiltonian allows for the following conserved quantity:
\begin{equation}
C=\sum_{m\neq0}\widetilde{c}^{\dagger}_m\widetilde{c}_m+a^{\dagger}a,
\end{equation}
which is easily checked to commute with the Hamiltonian. Next we compute the energy of the system for $C=1$ (noting that $C=0$ is trivial). The ground state (GS) in the absence of the cavity is spanned by the following states:
\begin{align}
|\psi^{\rm GS}_{0}\rangle&=|0_M\rangle\otimes|\underbrace{0\dots0}_{N-1}\rangle\,,\nonumber\\
|\psi^{\rm GS}_{1}\rangle&=|1_M\rangle\otimes|\underbrace{0\dots0}_{N-1}\rangle\,,
\end{align}
corresponding to the Majorana state being either empty ($0$) or occupied ($1$), while  all the other gapped states are empty. The excited states are then written as
\begin{eqnarray}
|\psi^{j}_0\rangle&=&|0_M\rangle\otimes|0\dots1_j\dots0\rangle\,,\nonumber\\
|\psi^{j}_1\rangle&=&|1_M\rangle\otimes|0\dots1_j\dots0\rangle\,,
\end{eqnarray}  
where $j=1,N-1$. We are now in position to write down the general state corresponding in the presence of the cavity to $C=1$ (one excitation) as
\begin{equation}
|\Phi_{ex}\rangle=\left[(a_0|0_M\rangle+a_1|1_M)\otimes|0\dots0\rangle\right]\otimes|1_{ph}\rangle+\left[\sum_{j=1}^{N-1}(b_0^j|0_M\rangle+b_1^j|1_M\rangle\otimes|0\dots1_j\dots0\rangle\right]\otimes|0_{ph}\rangle\,.
\end{equation}
We need to choose the coefficients $a_{0,1}$ and $b_{0,1}^j$ so that the state is an eigenstate of the total Hamiltonian:
\begin{equation}
(H_{\rm 1D}+H_{\rm int}^{\rm RWA}+\omega a^{\dagger}a)|\Phi_{\rm ex}\rangle=\epsilon|\Phi_{\rm ex}\rangle\,.
\end{equation}
First we  apply $H_{\rm int}^{\rm RWA}$ to the state $|\Phi_{\rm ex}\rangle$:
\begin{align}
H_{\rm int}^{\rm RWA}|\Phi_{\rm ex}\rangle=&\sum_{m,p,p'}S_m^{pp'}\left[(a_0|1_M\rangle+(-1)^{p'-1}a_1|0_M\rangle)\otimes|0\dots1_m\dots0\rangle\right]\otimes|0_{\rm ph}\rangle\nonumber\\
&+\sum_{m,p,p'}S_m^{pp'}(-1)^{p-1}\left[(b_0^m|1_M\rangle+(-1)^{p'-1}b_1^m|0_M\rangle)\otimes|0\dots0\rangle\right]\otimes|1_{\rm ph}\rangle\,.
\end{align}
On the other hand, the effect of $H_{\rm 1D}+H_{\rm ph}$ on this state reads:
\begin{align}
(H_{\rm 1D}+H_{\rm ph})|\Phi_{\rm ex}\rangle=&\left(\omega-\sum_{m\neq0}\epsilon_m\right)(a_0|0_M\rangle+a_1|1_M\rangle)\otimes|0\dots0\rangle\otimes|1_{ph}\rangle\nonumber\\
&-\left[\sum_{j;m\neq0}\epsilon_m(1-2\delta_{j,m})(b_0^j|0_M\rangle+b_1^j|1_M\rangle)|0\dots1_j\dots0\rangle\right]\otimes|0_{\rm ph}\rangle\,.
\end{align}

\begin{figure}[t]
\begin{center}
\includegraphics[width=0.9\linewidth]{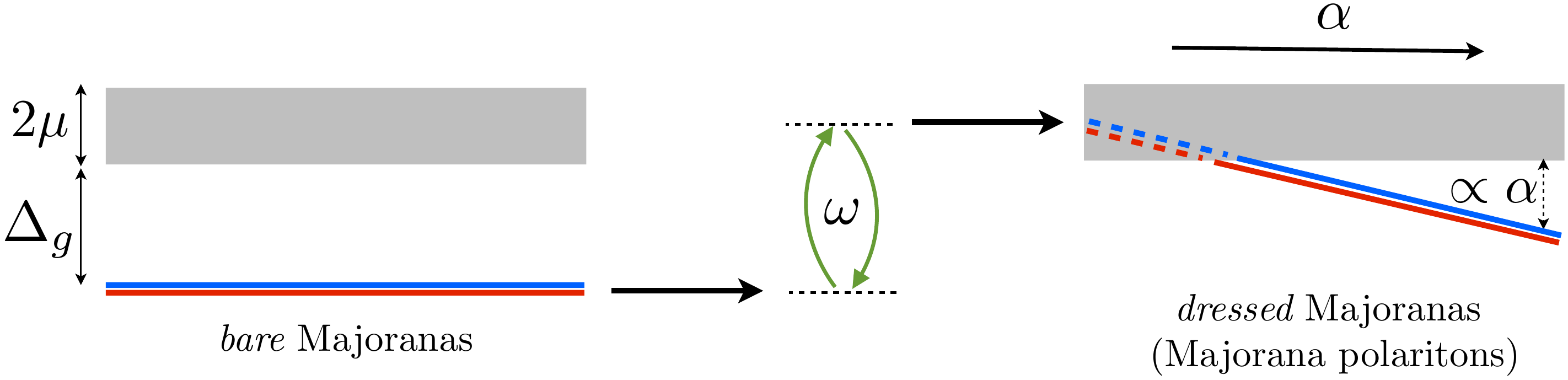}
\caption{Sketch of the energy-level structure for the one-excitation regime ($C=1$). The Majorana states in the absence of cavity (left) for parity $+$ (red) and $-$ (blue) are brought into the continuum via the interaction with the cavity. As $\alpha$ is increased, precisely one state for each parity is pushed below the continuum to form new Majoranas, now highly-entangled electron-photon states that can be called Majorana polaritons. Here, $\mu$ is the chemical potential, $\Delta_g$ stands for the initial gap from the zero mode to the continuum (in the absence of cavity), and $\omega$ is the cavity frequency.}
\label{sketch_energy_levels}
\end{center}
\end{figure}

We can now put everything together to find the equations for the energy: 
\begin{align}
\epsilon a_0&=(\omega-\sum_{m}\epsilon_m)a_0+\sum_{m,p,p'}(-1)^{p+p'}S_{m}^{pp'}b_1^m\,,\,\,\,\epsilon b_{1}^j=a_0\sum_{p,p'}S_j^{pp'}-b_1^j\sum_{m}\epsilon_m(1-2\delta_{j,m})\,,\nonumber\\
\epsilon a_1&=(\omega-\sum_{m}\epsilon_m)a_1+\sum_{m,p,p'}(-1)^{p-1}S_m^{pp'}b_0^m\,,\,\,\,\epsilon b_{0}^j=a_1\sum_{p,p'}(-1)^{p'-1}S_{j}^{pp'}-b_0^j\sum_{m}\epsilon_m(1-2\delta_{j,m})\,,
\end{align}
where it is evident that we are getting two decoupled sets of equations for each parity: $(a_0,\{b_1^m\})$ and $(a_1,\{b_0^m\})$. One can easily obtain from these relations the eigenenergy equations:
\begin{equation}
\left(\omega-\epsilon_{+(-)}-\sum_j\epsilon_j\right)+\sum_{m,r,r'}\sum_{p,p'}\frac{(-1)^{r+p'(r')}S_{m}^{rr'}S_m^{pp'}}{\epsilon_{+(-)}+\sum_{m'}\epsilon_{m'}(1-2\delta_{mm'})}=0\,,
\label{dicke_spec}
\end{equation}
for the parity  $p=+(-)$, which are quoted in the MT. Besides the spectrum $\epsilon$ we can also extract the set of coefficients $(a_{0(1)},\{b_{1(0)}^m\})$ (together with the normalization condition), which quantify the amount of photon-electron entanglement in the system. In the MT, we argued that the spectrum obtained by solving the Eq.~(\ref{dicke_spec}) yields one state for each parity emerging below the continuum of $C=1$ states for sufficiently large coupling strenght $\alpha$, which are degenerate (i.e., exponentially small splitting in the number of sites $N$). These states, which  we called Majorana polaritons, are  shown schematically in Fig.~\ref{sketch_energy_levels}. While for vanishingly small $\alpha$ the initially isolated Majoranas get mixed with the $C=1$ continuum of states, as $\alpha$ is increased further, new Majoranas reemerge, with well-defined parity and energy.

\end{widetext}

\end{document}